\documentclass[superscriptaddress,showpacs,preprintnumbers,amsmath,amssymb,twocolumn,prx]{revtex4-1}
\usepackage{graphicx}
\usepackage{verbatim} 
\usepackage[dvips]{color} 
\usepackage{epstopdf}

\usepackage{dcolumn}
\usepackage{bm}
\usepackage{upgreek}

\begin{document}


\title{High resolution two-dimensional optical spectroscopy of electron spins}

\author{M.~Salewski}
 	\affiliation{Experimentelle Physik 2, Technische Universit\"at Dortmund, 44221 Dortmund, Germany}
\author{S.~V.~Poltavtsev}
 	\affiliation{Experimentelle Physik 2, Technische Universit\"at Dortmund, 44221 Dortmund, Germany}
 	\affiliation{Spin Optics Laboratory, St. Petersburg State University, St. Petersburg 198504, Russia}
\author{I.~A.~Yugova}
    \affiliation{Spin Optics Laboratory, St. Petersburg State University, St. Petersburg 198504, Russia}
\author{G.~Karczewski}
 	\affiliation{Institute of Physics, Polish Academy of Sciences, PL-02668 Warsaw, Poland}
\author{M.~Wiater}
 	\affiliation{Institute of Physics, Polish Academy of Sciences, PL-02668 Warsaw, Poland}
\author{T.~Wojtowicz}
 	\affiliation{Institute of Physics, Polish Academy of Sciences, PL-02668 Warsaw, Poland}
    \affiliation{International Research Centre MagTop, PL-02668 Warsaw, Poland}
\author{D.~R.~Yakovlev}
\author{I.~A.~Akimov}
 	\affiliation{Experimentelle Physik 2, Technische Universit\"at Dortmund, 44221 Dortmund, Germany}
 	\affiliation{Ioffe Institute, Russian Academy of Sciences, 194021 St. Petersburg, Russia}
\author{T.~Meier}
    \affiliation{Department Physik \& CeOPP, Universit\"at Paderborn, D-33098 Paderborn, Germany}
\author{M.~Bayer}
 	\affiliation{Experimentelle Physik 2, Technische Universit\"at Dortmund, 44221 Dortmund, Germany}
 	\affiliation{Ioffe Institute, Russian Academy of Sciences, 194021 St. Petersburg, Russia}

\date{\today}


\begin{abstract}

Multidimensional coherent optical spectroscopy is one of the most powerful tools for investigating complex quantum mechanical systems. While it was conceived decades ago in magnetic resonance spectroscopy using micro- and radio-waves, it has recently been extended into the visible and UV spectral range. However, resolving MHz energy splittings with ultrashort laser pulses has still remained a challenge. Here, we analyze two-dimensional Fourier spectra for resonant optical excitation of resident electrons to localized trions or donor-bound excitons in semiconductor nanostructures subject to a transverse magnetic field. Particular attention is devoted to Raman coherence spectra which allow one to accurately evaluate tiny splittings of the electron ground state and to determine the relaxation times in the electron spin ensemble. A stimulated step-like Raman process induced by a sequence of two laser pulses creates a coherent superposition of the ground state doublet which can be retrieved only optically due to selective excitation of the same sub-ensemble with a third pulse. This provides the unique opportunity to distinguish between different complexes that are closely spaced in energy in an ensemble. The related experimental demonstration is based on photon echo measurements in an n-type CdTe/(Cd,Mg)Te quantum well structure detected by a heterodyne technique. The difference in the sub-$\mu$eV range between the Zeeman splittings of donor-bound electrons and electrons localized at potential fluctuations can be resolved even though the homogeneous linewidth of the optical transitions is larger by two orders of magnitude.

\end{abstract}


\maketitle

\section{Introduction}

Coherent optical spectroscopy has been widely used for the investigation of the energy levels of charge, vibration and spin states in condensed matter systems \cite{Spectroscopy-review,PhysToday2015}. It comprises many elaborated techniques that can be roughly divided in two main categories: the first set uses continuous wave light sources of tunable wavelength with narrow spectral width, i.e. long coherence time, which enable direct acquisition of coherent spectra in the frequency domain~\cite{EIT-Imamoglu}. The second set is based on recording the temporal dynamics of the system's response to short light pulse excitation~\cite{Shah-book}. Here, Fourier transformation of the optical transients into the frequency domain allows one to obtain similar information, not only about the energy level structure, but also about the inhomogeneous and homogeneous widths of the involved optical transitions as well as the coherent dynamics of the system under study. One of these techniques is the two-dimensional Fourier transform spectroscopy (2DFTS), which is based on the Fourier transformation of transient four-wave mixing (FWM) signals. This technique has rapidly developed during the last decade and has been successfully applied for the investigation of atomic, molecular, and condensed matter systems such as organic and inorganic semiconductors~\cite{Mukamel-book,Zhang-2007,Stone-2009,Mukamel-Cundiff-2007, Kasprzak-2011}. One of the appealing features of 2DFTS is the intuitive visualization of the underlying physics, as it enables one to extract not only energy levels but also provides a clear understanding of the dynamics and correlations between optical excitations~\cite{2DFS-review}.

Particular interest is attracted by systems with more than two states interacting with light. The most representative examples are $\rm V$- ($\Lambda$-) type energy level orders where the single ground (excited) state is optically coupled to an excited (ground) state doublet~\cite{Scully-book}. These level schemes make it possible to observe several fascinating phenomena such as quantum beats, coherent population trapping and electromagnetically induced transparency (EIT) which may be used for applications in quantum information technology~\cite{EIT-Imamoglu, Quantum-interface}. A main feature of a $\Lambda$-scheme is the possibly long coherence time of the ground states. In semiconductors, the $\Lambda$-scheme can be obtained for optical excitation of localized excess carriers, e.g., electrons in the conduction band or holes in the valence band, using their spin degree of freedom~\cite{Dyakonov-book}. In the case of resident electrons, the negatively charged exciton (trion $X^-$) and the donor-bound exciton ($D^0X$) are possible optically excited states. Coherent population trapping and EIT have been demonstrated for donor-bound excitons in bulk GaAs~\cite{Yamamoto-GaAs2005,Wal2010}, as well as trions in n-type CdTe quantum wells (QWs)~\cite{Wang2012} and quantum dots (QDs) charged with resident carriers~\cite{Steel2008, Warburton2009, Atature2014}. These studies have exploited high resolution spectroscopy with continuous wave lasers. Also spin control of resident carriers using ultrafast laser pulses has been successfully demonstrated~\cite{Greilich2006, Carter2007, Yamamoto2008}. However, most of the FWM and 2DFTS experiments on charged excitations in semiconductors have not exploited the spin degree of freedom in the ground state so far~\cite{Hoffmann1994, Tatarenko1999, Moody2013, Kasprzak2016}. Recently, we demonstrated that the ground state splitting of a trion in transverse magnetic field leads to quantum beats in the photon echoes at the Larmor precession frequency~\cite{Langer2012}. This allowed us to perform a coherent transfer of optical excitation into a spin ensemble and to observe long-lived photon echoes~\cite{Langer2014}. Here, we demonstrate that our approach can be used as a tool for spectroscopy of the ground state levels with remarkably high resolution: In an n-type CdTe/(Cd,Mg)Te QW we are able to resolve splittings between the spin sub-levels with sub-$\mu$eV precision and to distinguish between different types of electrons in the ensemble, namely electrons either bound to donors or localized on QW potential fluctuations. To that end, we show that stimulated step-like Raman processes in the two-pulse excitation scheme allow us to probe the electron spin ensemble with high selectivity and precision even for systems with broad optical transitions, given by large inhomogeneous broadening due to system variations or short optical coherence times leading to strong homogeneous broadening. Therefore our approach using ultrashort optical pulses mimics EIT which typically requires frequency stabilized lasers to resolve splittings of that magnitude.

The paper is organized as follows. First, we consider theoretically 2DFTS spectra under resonant excitation of localized trions or donor-bound excitons subject to a transverse magnetic field. We focus on Raman coherence spectra, which allow us to evaluate the splitting of the ground state and to determine the relaxation times in the electron spin ensemble. Second, we show experimental results on heterodyne-detected photon echoes recorded for a CdTe/(Cd,Mg)Te QW in which both trions and donor-bound exciton states are present.

\section{Two-dimensional Fourier spectra in nanostructures with resident electrons}

\begin{figure*}[htb]
\includegraphics[width=2\columnwidth]{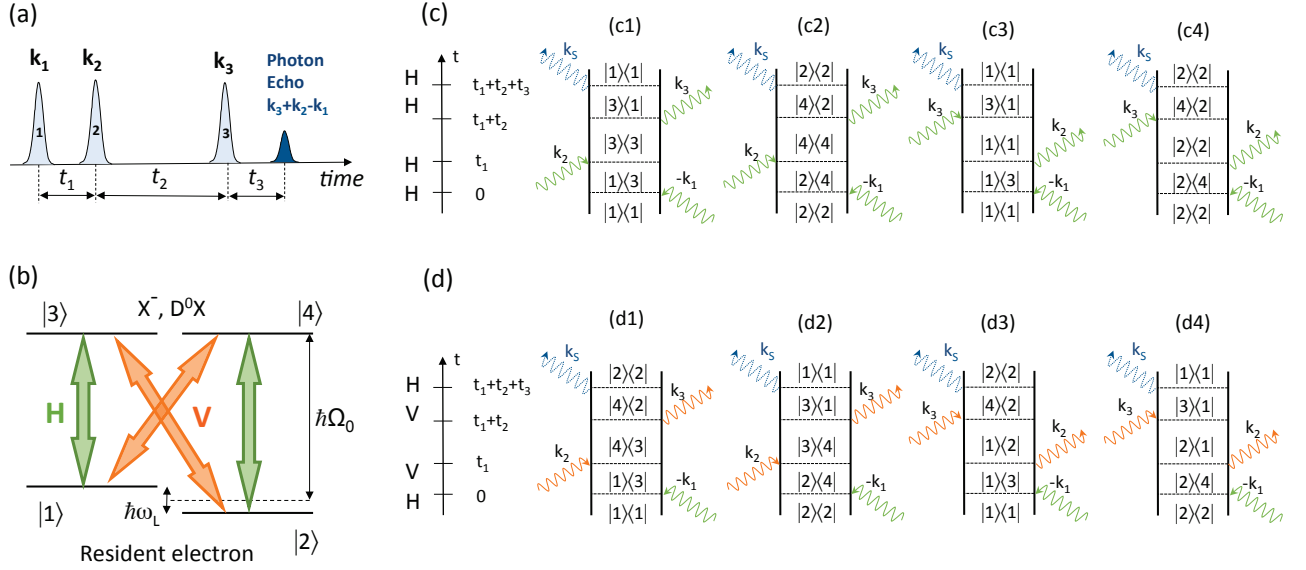}
\caption{(a) Scheme of the pulse sequence and the photon echo signal that is detected in $\mathbf{k}_S=\mathbf{k}_3+\mathbf{k}_2-\mathbf{k}_1$ direction. (b) Energy level diagram and optical transitions for the trion ($X^-$) or the donor-bound exciton ($D^0X$) which are localized in the semiconductor QW structure. The characters correspond to the polarization of the optical transition parallel (H) and perpendicular (V) to the magnetic field. (c) and (d) Double-sided Feynman diagrams for the HHH and HVV polarization configurations, respectively. For VVV and VHH the resulting diagrams are identical to HHH and HVV if the states $|3\rangle$ and $|4\rangle$ are exchanged. For the full set of diagrams see also section II of the supplementing material~\cite{supplement}.}
\label{fig:schema}
\end{figure*}

We concentrate on optical transitions in zincblende semiconductor nanostructures with resident electrons in the conduction band. In that way a $\Lambda$-transition scheme with long-lived coherence in the ground state can be established. We consider type-I semiconductor QWs with the heavy and light holes split by the confinement potential along the $z$-direction. At cryogenic temperatures the resident electrons are localized on potential fluctuations. Alternately the electrons can become trapped by donors. We consider an electron ensemble of low density where the electrons are well separated and do not interact with each other. The ground state is a doublet with electron spin $S=1/2$. The lowest energy, almost degenerate states that can be optically excited are the localized trion or the donor-bound exciton, whose angular momentum $J=3/2$ is given by the heavy hole. Higher energy states, e.g. the neutral exciton or the light hole exciton transitions, can be neglected if the optical pulses used for excitation are spectrally narrow enough. In a transverse magnetic field $B$ applied along the $x$-direction the electron spin states are split by $\hbar\omega_L=g\mu_B B$, where $\omega_L$ is the Larmor precession frequency, $g$ is the electron $g$ factor, and $\mu_B$ is the Bohr magneton. Optical transitions between all four states are allowed using light with linear polarization directed along (H) or perpendicular (V) to the direction of the magnetic field (H$\|x$ and V$\|y$).  The energy level structure and optical transitions are shown in Fig.~\ref{fig:schema}(b). The studied system can be considered as composed of two $\Lambda$ schemes sharing common ground states. This energy level structure is realized in a large variety of atomic and solid state objects with pseudospin in the ground and excited states~\cite{Scully-book}.

Transient FWM requires three optical pulses with wavevectors $\mathbf{k}_1$, $\mathbf{k}_2$ and $\mathbf{k}_3$, respectively, separated in time by $t_1$, $t_2$. The time $t_3$ gives the temporal delay of the resulting FWM signal with respect to pulse 3 (see Fig.~\ref{fig:schema}(a)). Due to the inhomogeneous broadening of the optical transitions the transient FWM signal is considered in the rephasing configuration and the resulting photon echoes are described by the optical field
\begin{equation}
\label{eq:PE-general}
S_{I}(t_1, t_2, t_3) = A_{\rm PE}(t_1, t_2,t_3) e^{ -\frac{(t_3-t_1)^2}{2\sigma^2}} e^{i\omega(t_3-t_1)},
\end{equation}
where $\sigma$ is the degree of (Gaussian) inhomogeneity and $\omega$ is the central frequency of the optical pulses which are tuned in resonance with the central frequency of the inhomogeneous ensemble~\cite{Shah-book, Mukamel-book}. The amplitude of the photon echo  $A_{\rm PE}(t_1, t_2, t_3)$ depends on the delay times $t_1$, $t_2$, and $t_3$ as well as on the polarization configuration of the exciting pulses. A proper polarization choice in the pulse sequence provides additional selectivity between various excitation paths~\cite{Langer2014}. This can be traced from the double-sided Feynman diagrams which are shown in Figs.~\ref{fig:schema}(c) and (d). The entirety of possible polarization configurations can be found in the supplementary material~\cite{supplement}. Here, we consider the HHH (VVV) and HVV (VHH) polarization sequences, which correspond to the most representative cases.  Following the Feynman diagrams it is seen that the resulting photon echo is H (V) polarized. For HHH the possible optical transitions take place between the two independent two-level pairs $|1\rangle-|3\rangle$ and $|2\rangle-|4\rangle$, while for HVV all transitions are involved and the coherent superposition between one pair of states is transferred to the other pair after each excitation event in a step-like process.

The first H-polarized pulse addresses the two optical transitions between $|1\rangle$ and $|3\rangle$ at frequency $\Omega_0 - \omega_L/2$ and between $|2\rangle$ and $|4\rangle$ at frequency $\Omega_0+\omega_L/2$. $\Omega_0$ corresponds to the optical resonance frequency at zero magnetic field ($\omega_L=0$). Pulse 1 creates two independent coherent superpositions between the pairs of states $|1\rangle - |3\rangle$  and $|2\rangle - |4\rangle$ which can be considered as {\it optical coherences}. In the density matrix formalism, they correspond to the $\rho_{13}$ and $\rho_{24}$ elements of the density matrix, respectively. Here, we assume that before excitation with pulse 1 the system is in the ground state and the only non-zero density matrix elements are $\rho_{11}=\rho_{22}=1/2$, i.e. $\hbar\omega_L\ll k_BT$, where $k_B$ is the Boltzman constant and $T$ is the temperature. The second and third pulses are both H- or V- polarized. Possible quantum mechanical pathways for the evolution of the system follow from the double-sided Feynman diagrams in Fig.~\ref{fig:schema} (c) and (d), respectively.

(i) {\it HHH co-polarized configuration}.  The second pulse addresses the same pairs of optical transitions and, in that way, the optical coherences $\rho_{13}$ and $\rho_{24}$ are transformed into the excited state populations $\rho_{33}$ and $\rho_{44}$ [diagrams (c1) and (c2) in Fig.~\ref{fig:schema}(c)] as well as the ground state populations $\rho_{11}$ and $\rho_{22}$ [diagrams (c3) and (c4) in Fig.~\ref{fig:schema}(c)]. The populations carry the information about the optical phase $\phi_{\pm} = (\Omega_0\pm\omega_L/2)t_1$ between the pulses 1 and 2, i.e. $\rho_{33}\propto\sin^2(\phi_-/2)$ and $\rho_{44}\propto\sin^2(\phi_+/2)$, where $\rho_{11}+\rho_{33}=\rho_{22}+\rho_{44}=1/2$ holds. The third pulse induces the coherences $\rho_{31}$ and $\rho_{42}$ and results in the emission of the photon echo after the rephasing process. At zero magnetic field the excited state populations are identical, $\rho_{33}=\rho_{44}$, and the dynamics are determined by the decay of the trion or donor-bound exciton complex when $t_2$ is varied. However, when the magnetic field is applied $\rho_{33}-\rho_{44}\propto\sin{(\omega_Lt_1/2)}\sin{(\Omega_0t_1)}$, i.e.  for a given $\Omega_0$ there are non-zero spin populations $J_x=(\rho_{33}-\rho_{44})/2$ and $S_x= (\rho_{11}-\rho_{22})/2 = -J_x$ in the excited and ground states, respectively. The spin populations carry the information about $\phi_{\pm}$ and correspondingly contribute to the coherent optical response~\cite{Langer2014}.

(ii) {\it HVV cross-polarized configuration}. Here, the second pulse accomplishes a stimulated step-like Raman process where the optical coherences $\rho_{13}$ and $\rho_{24}$ are transferred into the $X^-$ or $D^0X$ spin coherence $\rho_{34}$ [see the diagrams (d1) and (d2) in Fig.~\ref{fig:schema}(d)] and the electron spin coherence $\rho_{12}$ [see the diagrams (d3) and (d4) in Fig.~\ref{fig:schema}(d)]. The third pulse induces a back transformation of the trion and electron spin coherences into the optical coherences $\rho_{42}$ and $\rho_{31}$. This mechanism exploits off-diagonal density matrix elements. Thereby, the Raman process initiates a shift of the optical frequency of the emitted signal by $+\omega_L$ or $-\omega_L$ when starting from $\rho_{11}$ or $\rho_{22}$, respectively.

For the case that the splitting of the electron spin sub-levels in the ground state is smaller than the spectral width of the excitation laser pulses ($t_p\omega_L\ll2\pi$, where $t_p$ is the pulse duration) and the inhomogeneous broadening ($\omega_L\sigma\ll1$), the echo signals can be well approximated by Gaussian pulses with an amplitude $A_{\rm PE}(t_1,t_2)$ that depends on $t_1$ and $t_2$ only.
In the linear co-polarized polarization configuration
\begin{eqnarray}
\label{eq:A_PE-HHH}
A_{\rm PE}^{\|} &\propto& e^{-\frac{2t_1}{T_2}}e^{-\frac{t_2}{\tau_r}} \left[  2\cos^2(\omega_L t_1/2) + e^{-\frac{t_2}{T_h}} \sin^2(\omega_L t_1/2) \right] \nonumber \\
&+& e^{-\frac{2t_1}{T_2}}e^{-\frac{t_2}{T_1^e}}\sin^2(\omega_Lt_1/2),
\end{eqnarray}
where, $T_2$ and $\tau_r$ are the coherence time and the lifetime of the optically excited $X^-$ or $D^0X$ complex, $T_h$ is the spin relaxation time of the hole for $X^-$ or $D^0X$, and $T^e_1$ is the longitudinal spin relaxation time of the electron in the ground state~\cite{supplement}. The first term on the right hand side of Eq. \eqref{eq:A_PE-HHH} represents the population decay of the photoexcited complex (negative trion or donor-bound exciton) due to spontaneous recombination. The second and third terms originate from the spin relaxation of the hole in the excited state and the electron in the ground state, respectively. Note that the decay time of the last term is governed only by the spin relaxation time $T_1^e$ and represents the long-lived ground state spin population $S_x$ (for more details see also the supplementary material~\cite{supplement}). For $t_2=0$ the signal is:
\begin{equation}
\label{eq:A_PE-HHH-2pulse}
A_{\rm PE, t_2=0}^{\|} \propto e^{-\frac{2t_1}{T_2}},
\end{equation}
which is independent of the magnetic field.

In the cross-polarized linear polarization configuration, the signal is given by
\begin{eqnarray}
\label{eq:A_PE-HVV}
A_{\rm PE}^{\perp}  \propto  e^{-\frac{2t_1}{T_2}}   & &  \left[e^{-\frac{t_2}{\tau_T}} \cos(\omega_L t_1)
+ \right. \nonumber \\ & & \left. e^{-\frac{t_2}{T_2^e}} \cos(\omega_L (t_1 + t_2)) \right],
\end{eqnarray}
if we assume that $\omega_L  t_1 \gg 2\pi$, i.e. the Larmor precession is fast compared to the delay between the pulses 1 and 2~\cite{supplement}.
Here, $1/\tau_T$ is the inverse spin lifetime of the trion which is determined by the spin relaxation of the hole in the trion and its lifetime $1/\tau_T = 1/T_h + 1/\tau_r$. Similar to the previous case, we have two terms in Eq.~\eqref{eq:A_PE-HVV}, corresponding to two different contributions. The fast decay is attributed to the trion lifetime [diagrams (d1) and (d2) in Fig.~\ref{fig:schema}(d)] while the long-lived signal decays with $T_2^e$ [diagrams (d3) and (d4) in Fig.~\ref{fig:schema}(d)], which corresponds to the electron spin dephasing in the ground state (transversal spin relaxation time). For $t_2=0$ the signal transforms into
\begin{equation}
\label{eq:A_PE-HVV-2pulse}
A_{\rm PE, t_2=0}^{\perp} \propto e^{-\frac{2t_1}{T_2}}\cos(\omega_L t_1).
\end{equation}

\begin{figure}
\includegraphics[width=0.6 \columnwidth]{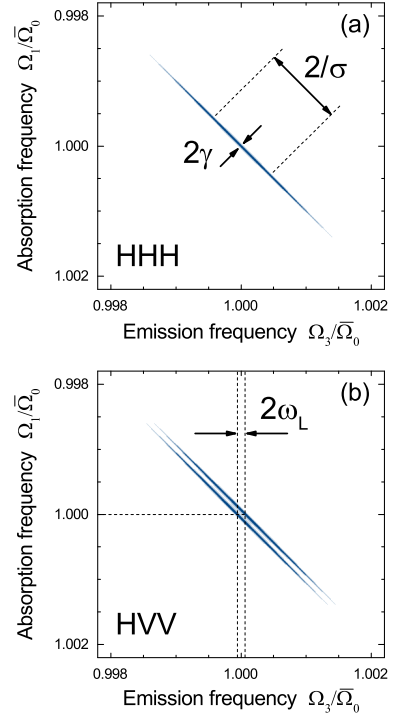}
\caption{Contour plots of the 2D Fourier rephasing spectra $S_I(\Omega_1,\Omega_3)$ in the co-polarized HHH (a) and in cross-polarized HVV (b) polarization configurations after Eqs.~\eqref{eq:rephasing-co-pol} and \eqref{eq:rephasing-cross-pol}, respectively. The following parameters are used in these calculations: $\hbar\overline{\Omega}_0=1.6$~eV, $\hbar\Gamma =\hbar/\sigma = 1$~meV, $\hbar\gamma=\hbar/T_2=10~\mu$eV,  $\hbar\omega_L = 100~\mu$eV.}
\label{fig:2DFS-rephasing}
\end{figure}

\begin{figure*}
\includegraphics[width=1.35\columnwidth]{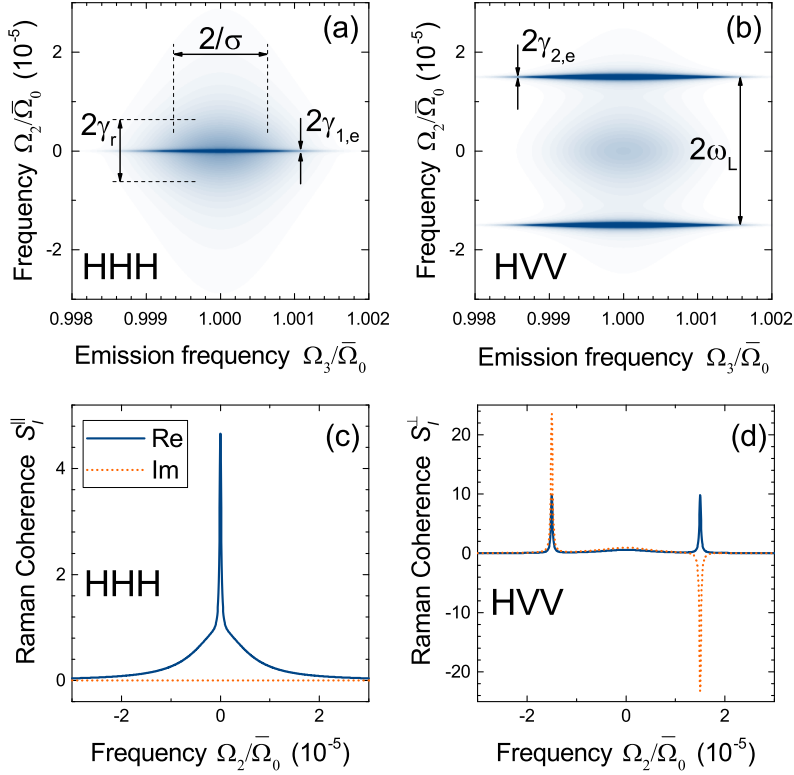}
\caption{Contour plots of the absolute value of the Raman coherence 2D Fourier spectra in the co-polarized HHH (a) and cross- polarized HVV (b) polarization configuration and their cross-sections at the optical frequency $\Omega_3 = \overline{\Omega}_0$ (c) and (d), respectively.  The following parameters are used: $\hbar\overline{\Omega}_0=1.6$~eV, $\hbar\Gamma =\hbar/\sigma = 1$~meV, $\hbar\gamma_r=\hbar\gamma_T=10~\mu$eV, $\hbar\gamma_{1,e}=\hbar\gamma_{2,e}=0.3~\mu$eV, $\hbar\omega_L = 24~\mu$eV and $t_1=26.7$~ps.}
\label{fig:2DFS-Raman}
\end{figure*}

Let us first consider the rephasing spectra for $t_2=0$
\begin{equation}
\label{eq:S_I}
S_{I}(\Omega_1,\Omega_3) = \int\int S_I(t_1,t_3) e^{i(\Omega_1t_1-\Omega_3t_3)} dt_1 dt_3,
\end{equation}
where $\Omega_1$ and $\Omega_3$ correspond to the absorption and emission optical frequencies. Fourier transformation of Eq.~\eqref{eq:PE-general} with Eqs. \eqref{eq:A_PE-HHH-2pulse} and \eqref{eq:A_PE-HVV-2pulse} gives:
\begin{eqnarray}
\label{eq:rephasing-co-pol}
S_{I}^{\|}(\Omega_1,\Omega_3) &\propto& \frac{\gamma e^{-\frac{\sigma^2(\Omega_1 -\overline{\Omega}_0)^2}{2}}}{4\gamma^2+(\Omega_1-\Omega_3)^2}, \\
\label{eq:rephasing-cross-pol} S_{I}^{\perp}(\Omega_1,\Omega_3) &=& S_{I}^+ + S_{I}^-, \\
S_{I}^\pm(\Omega_1,\Omega_3) &\propto& \frac{\gamma e^{-\frac{\sigma^2(\Omega_1 -\overline{\Omega}_0)^2}{2}}}{4\gamma^2+(\Omega_1-\Omega_3 \pm \omega_L)^2},
\end{eqnarray}
where $\overline{\Omega}_0$ is the central frequency of the inhomogeneous ensemble with a halfwidth in frequency corresponding to $1/\sigma$. Here, it is assumed that the inhomogeneous broadening $\Gamma=1/\sigma$ is significantly larger than the homogeneous broadening $\gamma=1/T_2$. This condition is fulfilled for the optical transitions to the localized $X^-$ and $D^0X$ complexes~\cite{Gobel1990,Ploog1993, Moody2013, Poltavtsev2017}. The rephasing spectra for the co- and cross-polarized configurations are plotted in Figs.~\ref{fig:2DFS-rephasing}(a) and (b), respectively. The splitting of the diagonal line by the energy $\hbar\omega_L$ in the cross-polarized configuration clearly demonstrates the stimulated Raman process with an increase ($S^+_I$) and a decrease ($S^-_I$) of the emission frequency compared to single excitation.

Particularly interesting are the Raman coherence spectra
\begin{equation}
\label{eq:S_IRaman}
S_I(t_1,\Omega_2,\Omega_3) = \int\int S_I(t_1,t_2,t_3) e^{-i(\Omega_2t_2+\Omega_3t_3)} dt_2 dt_3,
\end{equation}
where $\Omega_2$ is the Raman coherence frequency.

For the co-polarized configuration using Eqs.~\eqref{eq:PE-general}, \eqref{eq:A_PE-HHH}, and \eqref{eq:S_IRaman} we obtain
\begin{eqnarray}
\label{eq:HHH-Raman}
\nonumber &&S_I^{\|}(t_1,\Omega_2,\Omega_3) \propto e^{-\frac{2t_1}{T_2}}e^{-i \Omega_3 t_1}e^{-\frac{\sigma^2(\Omega_3-\overline{\Omega}_0)^2}{2}}(S_{I,T}^\|+S_{I,e}^\|), \\
\nonumber &&S_{I,T}^\| = 2\cos^2(\omega_L t_1/2) \frac{\gamma_r}{\gamma_r^2+\Omega_2^2} +\sin^2(\omega_L t_1/2)\frac{\gamma_T}{\gamma_T^2+\Omega_2^2},\\
&&S_{I,e}^\| = \sin^2(\omega_Lt_1/2)\frac{\gamma_{1,e}}{\gamma_{1,e}^2+\Omega_2^2},
\end{eqnarray}
where $\gamma_r=1/\tau_r$, $\gamma_T=1/\tau_T$ and $\gamma_{1,e}=1/T_1^e$. The corresponding 2DFTS image is shown in Fig.~\ref{fig:2DFS-Raman}(a) for the case when $\gamma_r = \gamma_T \gg \gamma_{1,e}$. In this case two Lorentzian peaks are centered at $\Omega_2=0$ (see the cross section at $\Omega_3=\overline{\Omega}_0$ in Fig.~\ref{fig:2DFS-Raman}(c)). Their widths are given by $\gamma_r$ and $\gamma_{1,e}$ and their relative amplitudes depend on $\omega_L t_1$. The spectrum can be used to evaluate the lifetimes of the excited states ($\gamma_r$) and the time of population relaxation between the ground states $|1\rangle$ and $|2\rangle$ ($\gamma_{1,e}$).

For the cross-polarized configuration using Eqs.~\eqref{eq:PE-general}, \eqref{eq:A_PE-HVV}, and \eqref{eq:S_IRaman} we obtain:
\begin{eqnarray}
\label{eq:HVV-Raman}
\nonumber && S_I^{\perp}(t_1,\Omega_2,\Omega_3) \propto e^{-\frac{2t_1}{T_2}}e^{-i \Omega_3 t_1}e^{-\frac{\sigma^2(\Omega_3-\overline{\Omega}_0)^2}{2}}(S_{I,T}^\perp+S_{I,e}^\perp), \\
\nonumber && S_{I,T}^\perp = 2\cos(\omega_L t_1)\frac{\gamma_T}{\gamma_T^2+\Omega_2^2},\\
&& S_{I,e}^\perp = \frac{\gamma_{2,e}e^{i\omega_L t_1}}{\gamma_{2,e}^2+(\Omega_2+\omega_L)^2}+ \frac{\gamma_{2,e}e^{-i\omega_L t_1}}{\gamma_{2,e}^2+(\Omega_2-\omega_L)^2},
\end{eqnarray}
where $\gamma_{2,e}=1/T_2^e$. This 2DFTS spectrum is shown in Fig.~\ref{fig:2DFS-Raman}(b). It is worth mentioning that in the HVV configuration only the imaginary part is present for $\omega_L t_1 = \pi/2$. It contains two peaks with different sign (dispersive shape) at frequencies $\Omega_2 = \pm \omega_L$ (see Fig.~\ref{fig:2DFS-Raman}(d)). The widths of these peaks are given by $\gamma_{2,e}$. Thus, the 2DFTS spectra measured in cross-polarized configuration allow us to evaluate the coherence times and the energy splitting between the ground state levels $\hbar\omega_L$. The  measurement of the splitting works even if it significantly undercuts the homogeneous spectral width of the optical transitions ($\gamma$). Thus, this method can be used for high resolution spectroscopy of the ground state. The advantage is the possibility to determine the splitting of the ground states for excitation at a particular photon energy $\omega=\Omega_1$.

An excellent example and demonstration of this powerful technique is the determination of the spin splittings of different complexes that exist simultaneously in the very same sample, e.g. $X^-$ and $D^0X$ as shown in the next section. This information cannot be obtained using pure spin resonance techniques where the optical excitation with $\Omega_1$ is absent. Therefore, in this particular case we perform optically detected magnetic resonance using coherent optical spectroscopy. Eventually the optical coherence initiated by the laser pulse plays an essential role during the excitation and the final emission process at $\Omega_1$ and $\Omega_3$ optical frequency, respectively. Otherwise, the Raman coherence cannot be restored which is in contrast to conventional time-resolved pump-probe Faraday rotation measurements~\cite{Crooker1997,Zhukov2007,Yugova2009}.

\begin{figure*}
\includegraphics[width=1.6\columnwidth]{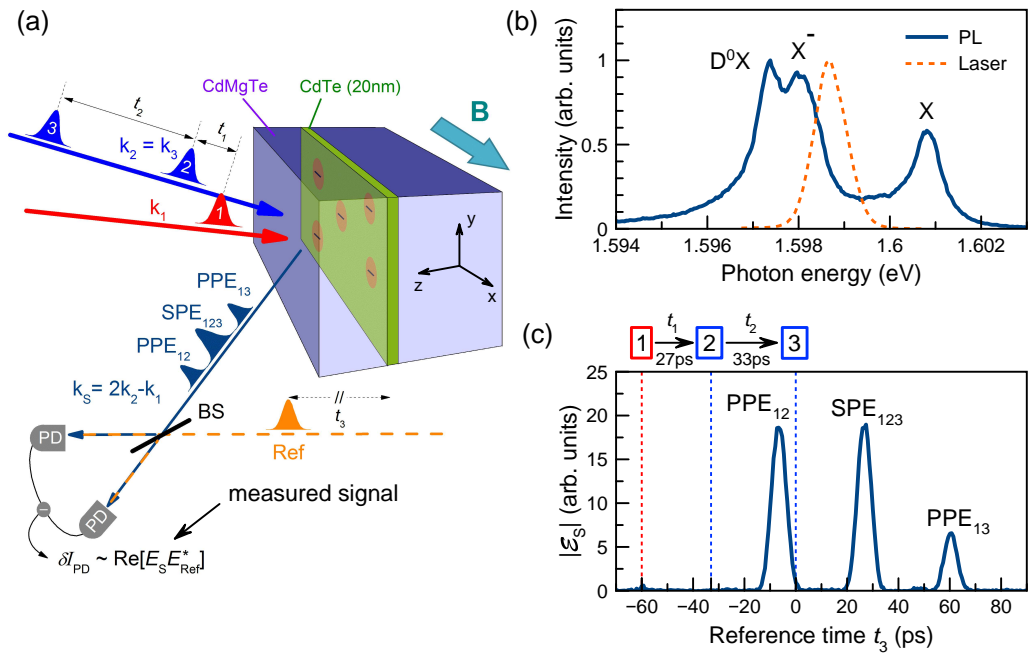}
\caption{(a) Schematic illustration of the heterodyne-detected photon echoes in reflection geometry. BS and PD denote beamsplitter and photodiode, respectively. (b) PL spectrum (the solid line) of the investigated 20$\,$nm thick CdTe/(Cd,Mg)Te QW structure measured at temperature $T=2\,$K for above-barrier excitation with photon energy $2.33\,$eV. The laser spectrum is shown by the dashed line. (c) Time-resolved cross-correlation of the resulting FWM signal $|\mathcal{E}_S(t_3)|$ measured for $t_1=27$~ps and $t_2=33$~ps as well as $\hbar\omega=1.597$~eV. The signal is given by photon echoes involving different pulse sequences: PPE$_{12}$ and PPE$_{13}$ correspond to the two-pulse sequences 1-2 and 1-3, respectively. SPE$_{123}$ corresponds to the three-pulse sequence 1-2-3.}
\label{fig:experiment}
\end{figure*}

\section{Optically detected coherent spectroscopy of ${\bf CdTe}$ QW with resident electrons}

The investigated sample comprises a 20-nm thick CdTe QW sandwiched between Cd$_{0.76}$Mg$_{0.24}$Te barriers. The QW was grown by molecular-beam epitaxy on a (100)-GaAs substrate, onto which a thick (Cd,Mg)Te buffer was deposited followed by a short period superlattice and a 100~nm Cd$_{0.76}$Mg$_{0.24}$Te barrier. The structure was not intentionally doped with donors; however, the unavoidable background of impurities results in localized resident carriers that originate from the barriers as well as from electrons bound to donors in the QW. The density of these electrons $n_e \leq 10^{10}~{\rm cm}^{-2}$ in the QW is low so that the exciton Bohr radius $a_B \ll 1/\sqrt{n_e}$, which allows us to consider each resident electron as isolated and non-interacting with other electrons. The photoluminescence (PL) spectrum is shown in Fig.~\ref{fig:experiment}(b) and consists of several spectral lines which we attribute to the exciton ($X$), the trion ($X^-$) and the donor-bound exciton ($D^0X$). In the experiments the sample is mounted in the Voigt geometry in a liquid-helium split-coil magneto-cryostat and is kept at a temperature of $T=2$~K. The direction of the magnetic field is parallel to the quantum well plane ($\mathbf{B} \parallel x$). The applied magnetic field $B=260$~mT results in an energy splitting of $\hbar\omega_L=24~\mu$eV which is small compared to the thermal energy $k_BT=170~\mu$eV.

For the photon echo experiments we use a sequence of three excitation pulses with variable delays $t_1$ between the pulses 1 and 2 and $t_2$ between the pulses 2 and 3 (see Fig.~\ref{fig:experiment}(a)). All pulses are obtained by splitting the emission of a tunable self-modelocked Ti-Sa oscillator with a repetition frequency of 75.75~MHz. The duration of the pulses is 2~ps and their spectral width is $\hbar\delta\omega=0.9$~meV. An example of the laser spectrum is shown in Fig.~\ref{fig:experiment}(b). These spectrally-narrow optical pulses allow us to achieve a high selectivity for excitation of the exciton complexes, e.g. $X$, $X^-$ or $D^0X$. In addition, they also prevent excitation of higher energy states, e.g. light hole transitions. The delay between the pulses is controlled using 27~cm long motorized translation stages by which we can cover maximum delays of about 1.8~ns. The optical pulse 1 hits the sample at an incidence angle of $6^\circ$. The pulses 2 and 3 hit both at  $7^\circ$ angle, so that their wavevectors are equal ($\mathbf{k}_2 = \mathbf{k}_3$). The beams are focused onto the sample at a spot with a diameter of about 200~$\mu$m. The energy density of each pulse is kept below 30~nJ/cm$^{2}$ in order to remain in the $\chi^{(3)}$-regime, i.e. the photon echo intensity depends linearly on the intensity of each of the beams. The polarization of the excitation pulses is controlled with retardation plates in conjunction with polarizers. The FWM signal is collected along the $\mathbf{k}_S =2\mathbf{k}_2-\mathbf{k}_1$ direction in reflection geometry. Here, the phase matching condition is not sensitive to the $z$-component of the wavevector $\mathbf{k}_S$ since the signal originates from the QW layer which has a thickness that is significantly smaller than the wavelength of light in CdTe.

In order to resolve the transient profile of the coherent optical response $E_S(t)$ we use interferometric heterodyne detection~\cite{Cho1992,QBits-book}, where the FWM signal and the reference beams are overlapped at the balanced detector (see Fig.~\ref{fig:experiment}(a)). The reference pulse with optical field $E_{\rm Ref}$ is obtained  from the same laser oscillator and its delay can be varied with a separate translation stage. The optical frequencies of pulse 1 and the reference pulse are shifted by $+40$~MHz and $-41$~MHz with acousto-optical modulators. The resulting interference signal ${\rm Re}[E_SE^*_{\rm Ref}]$ at the photodiode is detected with a high-frequency lock-in amplifier at $|2\omega_2 - \omega_1 - \omega_{\rm Ref}| = 1$~MHz. The phase of this signal is locked at short times because all of the pulses originate from the same laser source. However, random fluctuations of the optical phase in the different beam paths at time scales longer than 1 ms are not suppressed because no active stabilization of beam paths is implemented in our experiment. Therefore only the amplitude of the signal is accessible in the measurement procedure. Still, heterodyne detection provides a high-sensitivity, background-free measurement of the cross-correlation function for the absolute value of the FWM electric field amplitude $|\mathcal{E}_S(t_3)| \propto |\int E_S(t)E^*_{\rm Ref}(t-t_3)dt|$ when scanning the reference pulse delay time $t_3$, which is taken relative to the arrival time of pulse 3 (see Figs.~\ref{fig:schema} and \ref{fig:experiment}(a)).

\begin{figure}
\includegraphics[width=0.9\columnwidth]{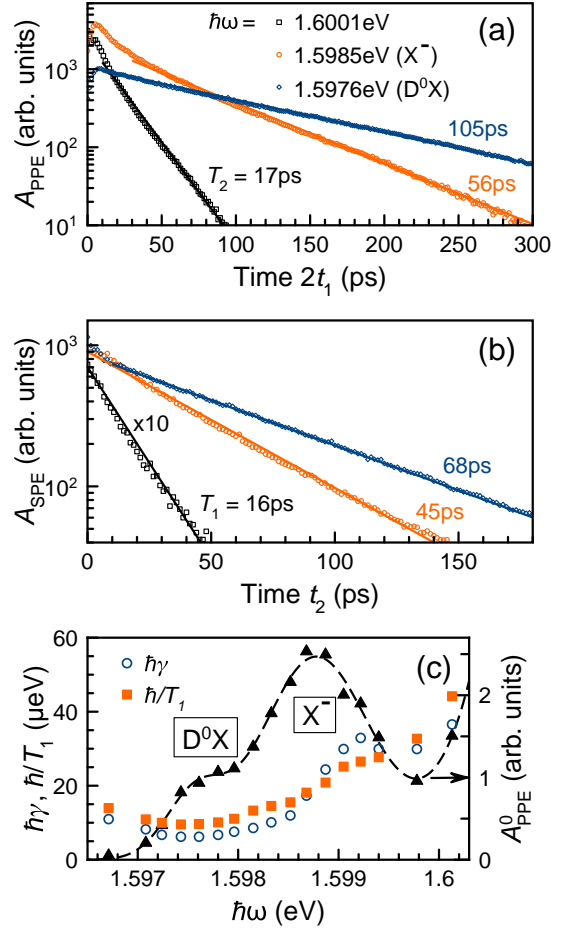}
\caption{Spectral dependences of the $A_{\rm PPE}$ and $A_{\rm SPE}$ transients at $B=0$. (a) and (b) $A_{\rm PPE}$ and $A_{\rm SPE}$ measured as function of $2t_1$ and $t_2$, respectively. The SPE is measured for $t_1=27$~ps. (c) Spectral dependencies of $A_{\rm PPE}^0=A_{\rm PPE}(t_1=0)$, $\hbar\gamma$ and $\hbar/T_1$, evaluated from the exponential fits to the echo transients for different photon energies $\hbar\omega$ (see the solid lines in (a) and (b))}
\label{fig:Times}
\end{figure}

Figure~\ref{fig:experiment}(c) shows a typical time-resolved FWM signal $|\mathcal{E}_S(t_3)|$ measured for $t_1=27$~ps and $t_2=33$~ps with excitation at photon energy $\hbar\omega=1.597$~eV. The coherent optical response is fully given by photon echoes. Due to $\mathbf{k}_2=\mathbf{k}_3$ several echoes are emitted along the phase matching direction $\mathbf{k}_S$. Two of them which appear at $t_3 = t_1 \pm t_2$ correspond to primary photon echoes (PPE) which result from the two-pulse sequences 1-2 and 1-3 and are labeled correspondingly as PPE$_{12}$ and PPE$_{13}$ in Fig.~\ref{fig:experiment}(c). The peak located at $t_3 = t_1$ corresponds to the stimulated photon echo (SPE) which is induced by the three-pulse sequence 1-2-3. Thus, the use of heterodyne detection allows us to distinguish between different echoes and to record the time evolution of the PE $A_{\rm PPE}$ and SPE $A_{\rm SPE}$ peak amplitudes by choosing the proper detection delay time $t_3$.

Figure \ref{fig:Times} summarizes the spectral dependence of the coherent optical response measured in the co-polarized configuration at $B=0$. The peak amplitudes $A_{\rm PPE}$ and $A_{\rm SPE}$ as function of $2t_1$ and $t_2$ for different $\hbar\omega$ are shown in Figs.~\ref{fig:Times}(a) and ~\ref{fig:Times}(b), respectively. From the exponential decay of the peak amplitudes we evaluate the coherence time $T_2$ and the lifetime $T_1$ of the photoexcited complexes following Eqs.~\eqref{eq:A_PE-HHH} and \eqref{eq:A_PE-HHH-2pulse}.
A proper choice of the excitation photon energy $\hbar\omega$ enables us to determine the homogeneous linewidth $\gamma$ and the population decay rate $\gamma_r=1/T_1$ of the particular optical transition, e.g. $X$, $X^-$ or $D^0X$. The spectral dependencies of the PPE signal strength $A^0_{\rm PPE}$, $\gamma$ and $\gamma_r$ are plotted in Fig.~\ref{fig:Times}(c).

\begin{figure*}
\includegraphics[width=1.6\columnwidth]{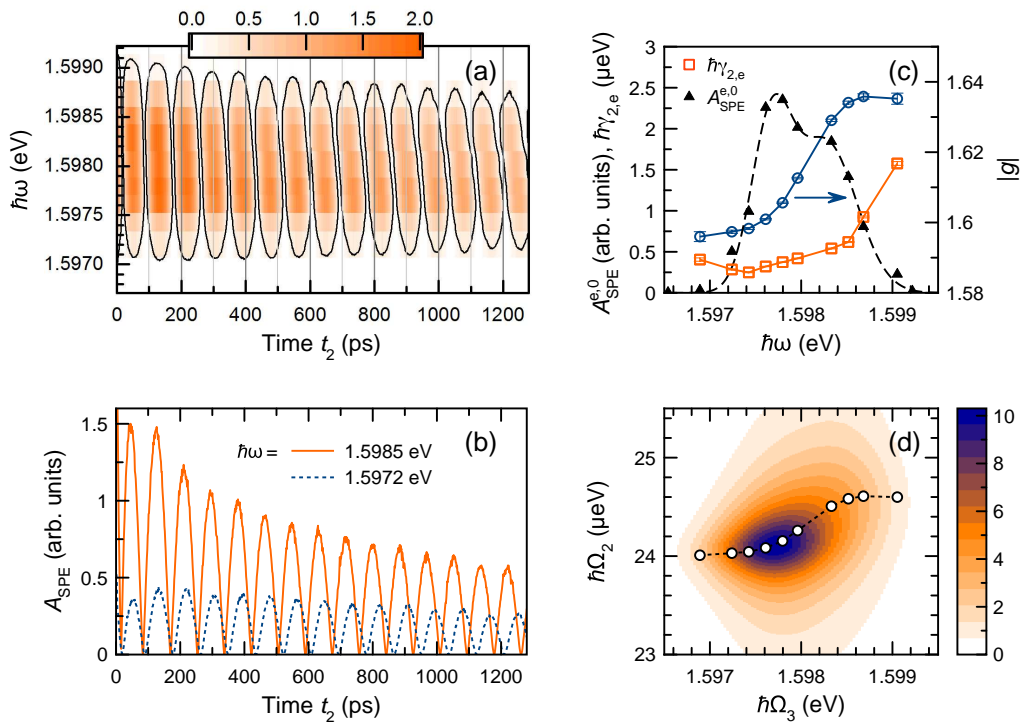}
\caption{(a) Contour plot of the long-lived SPE measured in the HVV polarization configuration as function of the excitation photon energy $\hbar\omega$ and delay time $t_2$. $B=260$~mT, $t_1=27$~ps.  The solid black lines indicate the signal level with 10\% of the maximum intensity. (b) Exemplary transients for given photon energies of 1.5972~eV and 1.5985~eV corresponding to optical excitation of the $D^0X$ and $X^-$ complexes, respectively. (c) Spectral dependence of the long-lived SPE signal strength $A^{e,0}_{\rm SPE}=A^e_{\rm SPE}(t_2=0)$ (left axis), the linewidth $\hbar\gamma_{2,e}$ (left axis) and the electron $g$ factor (right axis) evaluated from the $A_{\rm SPE}(t_2)$ transients at $B=260$~mT. (d) 2DFTS Raman coherence image $|S_{I}^\perp(\Omega_2,\Omega_3)|$ reconstructed from (c). The spectral dependence of $\hbar\omega_L$ is shown by the dots.}
\label{fig:2DFTS}
\end{figure*}

For resonant excitation of the low energy tail of the localized excitons at $\hbar\omega=1.601$~eV we obtain short decoherence times limited to 17~ps, i.e. $\hbar\gamma = 38~\mu$eV. For higher $\hbar\omega$ corresponding to excitation of free excitons ($X$) the linewidth becomes even larger reaching values of $150~\mu$eV~\cite{Langer2012,Moody2013}. For the trions and donor-bound excitons the homogeneous linewidths are significantly narrower due to the stronger localization of these complexes. Here, we obtain $\hbar\gamma \approx 12~\mu$eV for $X^-$ and $\hbar\gamma \approx 6~\mu$eV for $D^0X$.  At lower energies $\hbar\omega \leq 1.598$~eV we observe that the linewidth is determined mainly by the lifetime, i.e. $T_2 \sim 2T_1$ and thus, the pure dephasing is weak. Note, that the intentionally chosen spectrally narrow laser pulses ($\hbar\delta\omega=0.9$~meV) also help to suppress many-body interactions between different photoexcited complexes. For example, using spectrally-broad femtosecond pulses one would simultaneously excite excitons and trions. As a result, the exciton-trion interaction would lead to a significant increase of the homogeneous linewidth for the trion transition, which would become comparable to the one measured for the $X$~\cite{Moody2013}.

In order to evaluate the Zeeman energy splitting of the ground electronic states we concentrate on the $A_{\rm SPE}$ transients setting the reference pulse delay $t_3=t_1$ and performing scans as function of $t_2$ for different excitation photon energies $\hbar\omega$ in cross-polarized configuration (HVV). The data are summarized in Fig.~\ref{fig:2DFTS}. Contour plots of the SPE peak amplitude as function of $t_2$ and $\hbar\omega$ measured at $B=260$~mT for $t_1=27$~ps are shown in Fig.~\ref{fig:2DFTS}(a). Exemplary curves taken at $\hbar\omega=1.5972$~eV and 1.5985~eV, corresponding to excitation of the $D^0X$ and $X^-$ optical transitions, respectively, are shown in Fig.~\ref{fig:2DFTS}(b). When $t_2$ is varied we observe an oscillatory signal that decays on a long time-scale of several ns. Note, that the long-lived signal is present only in the spectral region $1.597-1.599$~eV where the $X^-$ and the $D^0X$ resonances are located, i.e. it is necessary to address resident carriers for storing optical information on such long time scales.

Using Eq.~\eqref{eq:A_PE-HVV} we evaluate the spectral dependences of the long-lived SPE signal strength $A^{e,0}_{\rm SPE}$, the absolute value of the electron $g$ factor $|g|=\hbar\omega_L/\mu_BB$ and the decay rate $\gamma_{2,e}$ which are plotted in Fig.~\ref{fig:2DFTS}(c). In accordance with Eq.~\eqref{eq:HVV-Raman} these parameters allow us to restore the 2DFTS Raman coherence image $|S_{I}^\perp(\Omega_2,\Omega_3)| \approx \sum A^{e,0}_{\rm SPE} |S^\perp_{I,e}(\Omega_2)|\exp{(-\frac{\sigma^2(\Omega_3-\omega)^2}{2})}$, where we set $2/\sigma = \delta\omega$ and neglect the $S^\perp_{I,T}$ term since $\gamma_T \gg \gamma_{2,e}$. The sum is taken over all excitation energies $\hbar\omega$ used in the experiment.  The resulting contour plot is shown in Fig.~\ref{fig:2DFTS}(d). Note that we cannot distinguish between real and imaginary contributions to the 2DFTS signal. However, as follows from the theoretical description in Sec.~II, this is not crucial for the evaluation of the energy splitting in the ground state in case of isolated localized electrons. Phase stabilized measurements have, however, great potential for in-depth studies of many-body interactions in an ensemble of resident electrons similar to that obtained for photoexcited exciton complexes~\cite{2DFS-review}.

The most striking feature of Fig.~\ref{fig:2DFTS}(d) is the variation of the Raman coherence peak with photon energy $\hbar\Omega_3$ demonstrating a step-like behavior and clearly showing that $\hbar\omega_L$ increases from $24.0$ to $24.6~\mu$eV when the excitation energy is varied from $D^0X$ to $X^-$. The extracted $g$ factors are $|g|=1.595$ and 1.635 for the resident electrons bound to a donor and localized in a potential fluctuation, respectively (see also Fig.~\ref{fig:2DFTS}(c)). The small difference between these values have the opposite trend to that expected for free electrons in CdTe/(Cd,Mg)Te QWs and (Cd,Mg)Te \cite{Sirenko1997}. The electron $g$ factor is controlled by the admixture of valence band states to the conduction band, which in turn is dependent not only on the band gap energy, but also on the electron and hole localization. This localization is different for the $D^0X$ and $X^-$ complexes, which results in the measured differences of the $g$ values.

Finally, we discuss the spectral dependence of the decay rate $\gamma_{2,e}$, which increases from $0.25~\mu$eV for donor bound electrons to $1.5~\mu$eV for high energy trions. There are several mechanisms which can contribute to the decay of the long-lived signal. For localized non-interacting electrons spin relaxation and hopping between localization sites are relevant. The latter mechanism deserves special attention as it may be spin-conserving and, therefore, does not give a contribution to conventional pump-probe measurements where the decay of the signal is determined solely by the spin relaxation. The increase of $\gamma_{2,e}$ with increasing $\hbar\omega$ indicates that hoping of electrons between localization sites plays an important role for states with weaker localization. Simultaneously, for the donor bound electrons with strongest localization the signal decay approaches the values measured on similar structures using the pump-probe technique ($\sim 3-10$~ns)~\cite{Zhukov2007}.

Thus our results demonstrate that the decay of Raman coherence in an electron spin ensemble measured by photon echoes provides access to the local relaxation processes, such as hoping of carriers between localization sites or spin interactions between electrons within the ensemble (e.g. spin flip-flops). This is because every individual electron in the ground state contributes to the coherent optical response only if it is addressed by all three optical pulses sequentially. Moreover in photon echo experiments due to dephasing of optically excited states between the first and second optical pulses no macroscopic spin polarization is created in the ground state after the stimulated Raman process. In contrast, time resolved pump probe Faraday rotation \cite{Crooker1997,Zhukov2007,Yugova2009} and transient spin gratings techniques \cite{Cameron-1996, Cundiff-2006, Marie-2013} detect the evolution of the macroscopic spin polarization for a large electron ensemble and local relaxation processes cannot be probed directly. Therefore a comparison of the signal decays recorded with different experimental techniques can be used to obtain the full and self-consistent physical picture of the spin dynamics in electron ensembles.
\\

\section{Conclusions}

In conclusion, we have demonstrated that two-dimensional Fourier spectroscopy addressing photon echoes can be successfully applied for the evaluation of tiny splittings between ground state energy levels which are optically coupled to a common excited state in, e.g., a $\Lambda$-type scheme. We have shown that the stimulated step-like Raman process induced by the sequence of two pulses creates a coherent superposition of the ground state doublet which can be retrieved only by optical means due to selective excitation of the same spin sub-ensemble with the third pulse. This provides the unique opportunity to distinguish between several electron spin species in a large ensemble of emitters. As a proof of principle we have applied this method to an n-type CdTe/(Cd,Mg)Te quantum well system for which the Zeeman splitting difference in the sub-$\mu$eV range between donor-bound electrons and electrons localized on potential fluctuations has been resolved while the homogeneous linewidth of the optical transitions is two orders of magnitude larger than this splitting. Our results pave the way for further developing two-dimensional Fourier imaging into a high resolution spectroscopy tool, independent on the nature of the energy splitting in the ground state.

\section{Acknowledgements}

We acknowledge the financial support by the Deutsche Forschungsgemeinschaft (DFG) through the Collaborative Research Centre TRR 142 (project A02) and the International Collaborative Research Centre TRR 160, the latter of which is also supported by the Russian Foundation of Basic Research (project N 15-52-12016 NNIO\_a). The research in Poland was partially supported by the National Science Centre (Poland) through Grants No. DEC-2012/06/A/ST3/00247 and No. DEC- 2014/14/M/ST3/00484, as well as by the Foundation for Polish Science through the IRA Programme co-financed by EU within SG OP.

\end{document}